# Magnetic nonreciprocity in a hybrid device of asymmetric artificial spin-ice-superconductors

Chong Li(李冲)[1,2,3], Peiyuan Huang(黄培源)[1,2,3], Chen-Guang Wang(王晨光)[1,2,3], Haojie Li(李浩杰)[1], Yang-Yang Lyu(吕阳阳)[1,**], Wen-Cheng Yue(岳文诚)[1,**], Zixiong Yuan(袁子雄)[1,2,3], Tianyu Li(李甜雨)[1,2,3], Xuecou Tu(涂学凑)[1,2], Tao Tao(陶涛)[3], Sining Dong(董思宁)[1,3], Liang He(何亮)[3], Xiaoqing Jia(贾小氢)[1], Guozhu Sun(孙国柱)[1,2], Lin Kang(康琳)[1], Huabing Wang(王华兵)[1,2], Peiheng Wu(吴培亨)[1,2], and Yong-Lei Wang(王永磊)[1,2,3,**]

[1]Research Institute of Superconductor Electronics, Nanjing University, Nanjing, China
[2]Purple Mountain Laboratories, Nanjing, China
[3]State Key Laboratory of Spintronics Devices and Technologies, Nanjing University, Suzhou, China

[*]This work is supported by the National Natural Science Foundation of China (62288101 and 62274086)，the National Key R&D Program of China (2021YFA0718802), Jiangsu Outstanding Postdoctoral Program.
[**]Corresponding authors. Email: yylyu@nju.edu.cn; wenchengyue@nju.edu.cn; yongleiwang@nju.edu.cn;

Controlling the size and distribution of potential barriers within a medium of interacting particles can unveil unique collective behaviors and innovative functionalities. In this study, we introduce a unique superconducting hybrid device using a novel artificial spin ice structure composed of asymmetric nanomagnets. This structure forms a distinctive superconducting pinning potential that steers unconventional motion of superconducting vortices, thereby inducing a magnetic nonreciprocal effect, in contrast to the electric nonreciprocal effect commonly observed in superconducting diodes. Furthermore, the polarity of the magnetic nonreciprocity is in-situ reversible through the tunable magnetic patterns of artificial spin ice. Our findings demonstrate that artificial spin ice not only precisely modulates superconducting characteristics but also opens the door to novel functionalities, offering a groundbreaking paradigm for superconducting electronics.

In type-II superconductors, magnetic fluxes penetrating the material are quantized and referred to as vortices; each vortex has a quantized flux of $\Phi_0 = h/2e = 2.068 \times 10^{-15}$ Wb and consists of supercurrents encircling non-superconducting cores. The vortices are propelled by the Lorentz force when an external current is applied, and their motion generates resistance and dissipates energy. Consequently, the electromagnetic properties of superconducting materials are dominated by the motion behavior of these vortices. Controlling vortex motion is crucial for the development of innovative electronic devices, such as vortex pumps, diodes, and rectifiers.[1, 2] Vortex dynamics has become an active research area, both in fundamental studies of vortex matter and in applications in electronics.[3] Previous research has demonstrated that vortices can be trapped by nanoscale defects, including those created by irradiation,[4, 5] grain boundaries,[6-8] doping[9] and nanostructures fabricated using nanofabrication techniques.[10-17] However, once the samples are prepared using these methods, the spatial distribution of the vortices' pinning potential landscape is fixed. To introduce flexibility in tailoring superconducting vortices, earlier studies have indicated that superconducting vortices can interact with magnetic nanostructures.[18-28] By applying an external magnetic field, the magnetization of these structures can be controlled in situ, which alters the pinning potential distribution and enables reconfigurability of superconductivity.[26-28]

Recently, a magnetic structure known as artificial spin ice (ASI) has been utilized as a reconfigurable pinning potential to control vortices in superconducting heterostructures.[29-31] ASI, an array of nanomagnets, allows for the alteration of the magnetization direction of these nanomagnets through an in-plane external magnetic field. This enables ASI to create a variety of magnetic charge structures, offering diverse regulatory effects on the vortices and thus allowing for flexible customization of superconducting properties and functionalities. Examples include switchable geometric vortex frustration,[30] reversible vortex Hall effect,[31] and programmable vortex diodes.[29, 30] However, prior research on vortex control using ASI all relied on symmetrical nanomagnets. In this paper, we introduce a novel square ASI structure made of asymmetric nanomagnets. The unique shape of these nanomagnets induces a unique magnetic non-reciprocal effect, which can be reconfigured in situ by altering the magnetization states of the ASI. Analogous to the electric nonreciprocal effect in superconducting diodes, the magnetic nonreciprocity acts like a magnetic-field-driven superconducting 'diode', wherein zero-resistance is observed only when the magnetic field is applied in one specific direction. Our findings underscore the pivotal role of locally asymmetric magnetic potential in unlocking distinctive superconducting properties, setting the stage for a new paradigm in superconducting electronics.

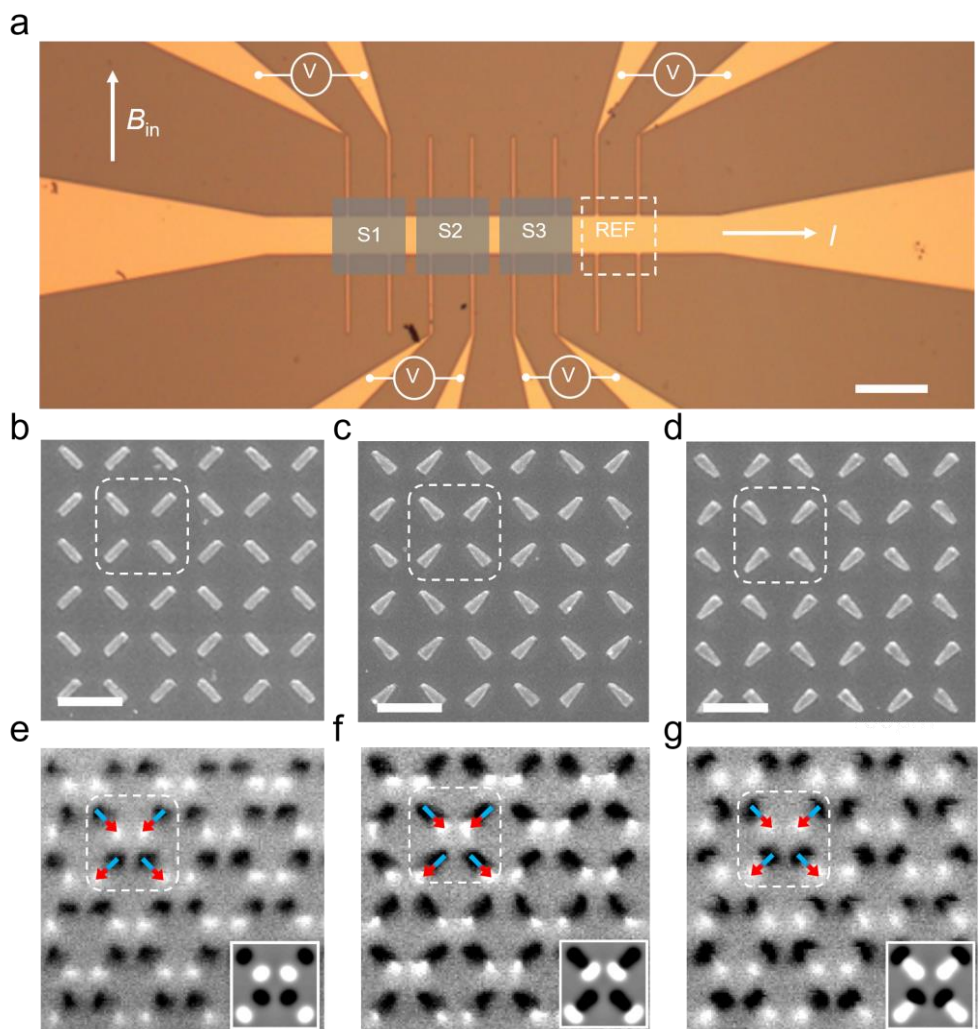

**Fig. 1. Device and sample structure.** a, Optical image of a superconducting MoGe microbridge with four sections. Scale bar, 100 μm. b-d, Scanning electron microscopy (SEM) images showing square-ASIs structures in Sections S1-S3, respectively. Scale bar: 700 nm. The dashed boxes highlight a magnetic vertex containing four nanobars. e-g, Magnetic force microscopy (MFM) images of the square-ASI structures in (b)-(d), respectively. The white/black spots reveal positive/negative magnetic charges, while the arrows indicate the magnetizations of single-domain nanomagnets.

Our experiment was conducted using a superconductor-ASI heterogeneous structure. An optical image of the sample is presented in Figure 1a. The device comprises a 50 nm thick $Mo_{79}Ge_{21}$(MoGe) film microbridge, which was deposited on a silicon wafer employing standard photolithography and magnetron sputtering techniques. The ASI structure was subsequently fabricated on top of the MoGe film using e-beam lithography and e-beam evaporation. Detailed methodologies of the sample fabrication are documented elsewhere.[30, 31] As depicted in Figure 1a, the sample is segmented into three sections (S1-S3), each featuring a unique ASI structure, alongside a section (REF) that retains a bare MoGe film serving as a reference. Figures 1b-1d showcase scanning electron microscopy images of three distinct ASI structures. Section S1 consists of an array of symmetrical nanomagnets, with dimensions of 300 nm (length) x 80 nm (width) x 25 nm (thickness). The ASI array in Section S2 comprises asymmetrically trapezoidal nanomagnets, each sized at 300 nm (altitude) x 40 nm (base1) x 120 nm (base2) x 25 nm (thickness). Conversely, the ASI array in Section S3 features geometrically inverted nanomagnets relative to those in Section S2.

The interaction between the ASI and the superconducting film is dominated by the magnetic charges of ASI's nanomagnets and superconducting vortices. Each nanomagnet can be considered as a dumbbell consisting of a pair of positive and negative magnetic charges, which respectively repel (attract) and attract (repel) vortices depending on the directions of magnetic fields.[30, 31] Therefore, the distribution of the magnetic charge patterns is crucial for controlling vortex dynamics in superconductors. The magnetic charge patterns of ASIs can be directly revealed from magnetic force microscopy (MFM). Figures 1e-1g show MFM images of the ASI magnetized using an in-plane magnetic field pointing to the bottom. The magnetization directions of nanomagnets are indicated by arrows in Figures 1e-1g. The white and black spots display the positive and negative magnetic charges, respectively. The positive and negative magnetic charges in Section S1 are equal in size and strength (Figure 1e), while Section S2 displays smaller positive charges and larger negative charges (Figure 1f), and Section S3 reveals inversed geometries (as compared to Section S2) of positive and negative charges (Figure 1g). The geometric structures of the magnetic charge patterns can be revealed more clearly from micromagnetic simulations in the insets of Figures 1e-1g.

To investigate the impact of the magnetic charge potentials on the motion of sup-erconducting vortices, we conducted magnetotransport experiments in a three-axis superconducting vector magnet. This apparatus can generate a magnetic field in any desired 3D orientation, enabling in-situ control over the magnetic charge potentials and the density of superconducting vortices. The in-plane magnetic field is utilized solely to magnetize the ASI, establishing specific magnetic charge potentials, and is removed during transport measurements. An out-of-plane magnetic field B is applied to adjust the density and polarities of vortices within the MoGe film. The superconducting transition temperature for the MoGe film is identified at 6.6 K (Figure S3), with the transport experiments conducted at 5.7 K.

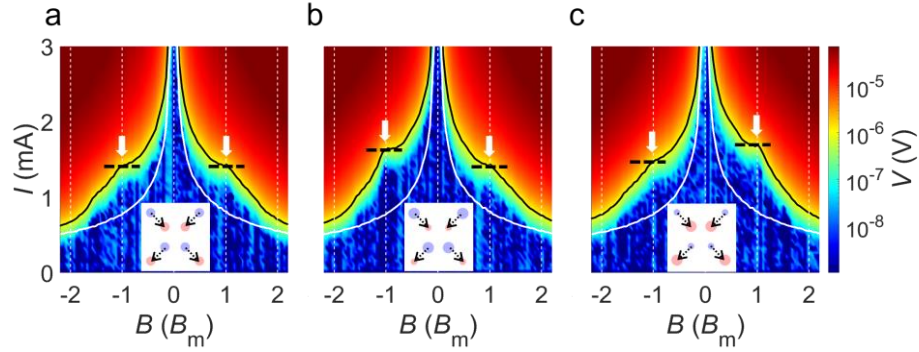

**Fig. 2. Magnetic nonreciprocal effect.** a-c, Dissipation voltage induced by vortices/antivortices motion, dependent on both current and magnetic field, the three maps correspond to the results obtained under magnetic charge states of Figures 1e-g, respectively. The magnetic field unit is normalized by the matching field $B_m$ = 78.9 Gs. Black lines represent critical current curves extracted with a voltage criterion of 1 µV, and white lines represent critical current curves in the REF section. The red/blue spots reveal positive/negative magnetic charges in insets, while the arrows indicate the magnetizations of single-domain nanomagnets.

Figures 2a-2c illustrate the magnetic field and current dependences of the dissipation voltages measured under the magnetic charge patterns in Figures 1e-1g, respectively. The magnetic field values are normalized by the first matching field $B_m$ = 78.9 Gs, at which the vortex density equals the nanomagnet density. The insets in Figures 2a-2c depict schematic diagrams of the magnetic charge patterns, correlating with Figures 1e-1g, respectively. The black solid lines in Figures 2a-2c represent the critical current ($I_c$) curves, determined using a voltage criterion of 1 µV. The white solid line in each figure corresponds to the $I_c$ curve of section REF for the bare MoGe film. These results indicate that the ASI structures significantly enhance $I_c$, presenting a strong vortex pinning effect from the magnetic charges. Additionally, the $I_c$ curves of sections S1-S3 with ASIs exhibit clear peaks/kinks at $B = \pm B_m$ (Figures 2a-2c), which is a typical feature of the vortex matching effect.[30, 31] The $I_c$ curve of Section S1, employing symmetric nanomagnets, shows a symmetric response with respect to magnetic field $B$ (Figure 2a), similar to the symmetrical effects observed in hybrid devices of square-ASI/superconductor[30] and pinwheel-ASI/superconductor.[31] Figure 2b highlights a significant finding that the $I_c$ curve is asymmetrical relative to $B = 0$. Specifically, the critical currents under negative fields surpass those under positive fields, implying that vortices move more easily at positive fields than at negative fields. This phenomenon leads to a magnetic nonreciprocal effect triggered by asymmetric nanomagnets. Interestingly, Figure 2c demonstrates that the polarity of the magnetic nonreciprocity can be reversed by inverting the asymmetric geometry of the nanomagnets. These magnetic nonreciprocal effects can also be repeated in another device (See Supplementary Fig. S1).

To elucidate the microscopic mechanism behind the magnetic nonreciprocal effect and its association with the asymmetry of ASI's nanomagnets, we performed molecular dynamic (MD) simulations on our devices. The detailed MD simulation methodology is described elsewhere.[30, 31] Figures 3a–3c display the simulated $I_c$ curves corresponding to Figures 2a–2c, respectively. They show a nice consistency between the experimental outcomes and simulations in terms of symmetric and/or asymmetric responses across all sample structures. Except for clear matching effects at $B = \pm B_\mathrm{m}$, figures 3a-3c also show matching peaks at $B = \pm 1.5 B_\mathrm{m}$, which is due to collective motion of interstitial vertices (Supplementary Fig. S2). Very weak kinks at $B = \pm 1.5 B_\mathrm{m}$ can also be observed in experiments (Figs. 2a-2c). To further unveil the microscopic nature of magnetic symmetry/asymmetry, we simulated the vortex motion behavior under positive and negative fields, which are presented in Videos 1-3. Screenshots depicting the vortices' trajectories are shown in Figures 3d-3i. Given that the driving force on a vortex from the applied current is opposite to that on an antivortex (which possesses inverted magnetic flux polarity and reversed circulating supercurrent) from the same current, vortices (under a positive magnetic field) and antivortices (under a negative magnetic field) move in opposite directions. For Section S1 with symmetric nanomagnets, the size and field strength of the positive and negative magnetic charges are identical, except for their polarity. Consequently, vortices/antivortices exhibit equivalent motion (Video 1, Figures 3d and 3g) under positive and negative fields, though in reversed directions, leading to a symmetric transport response (Figure 3a).

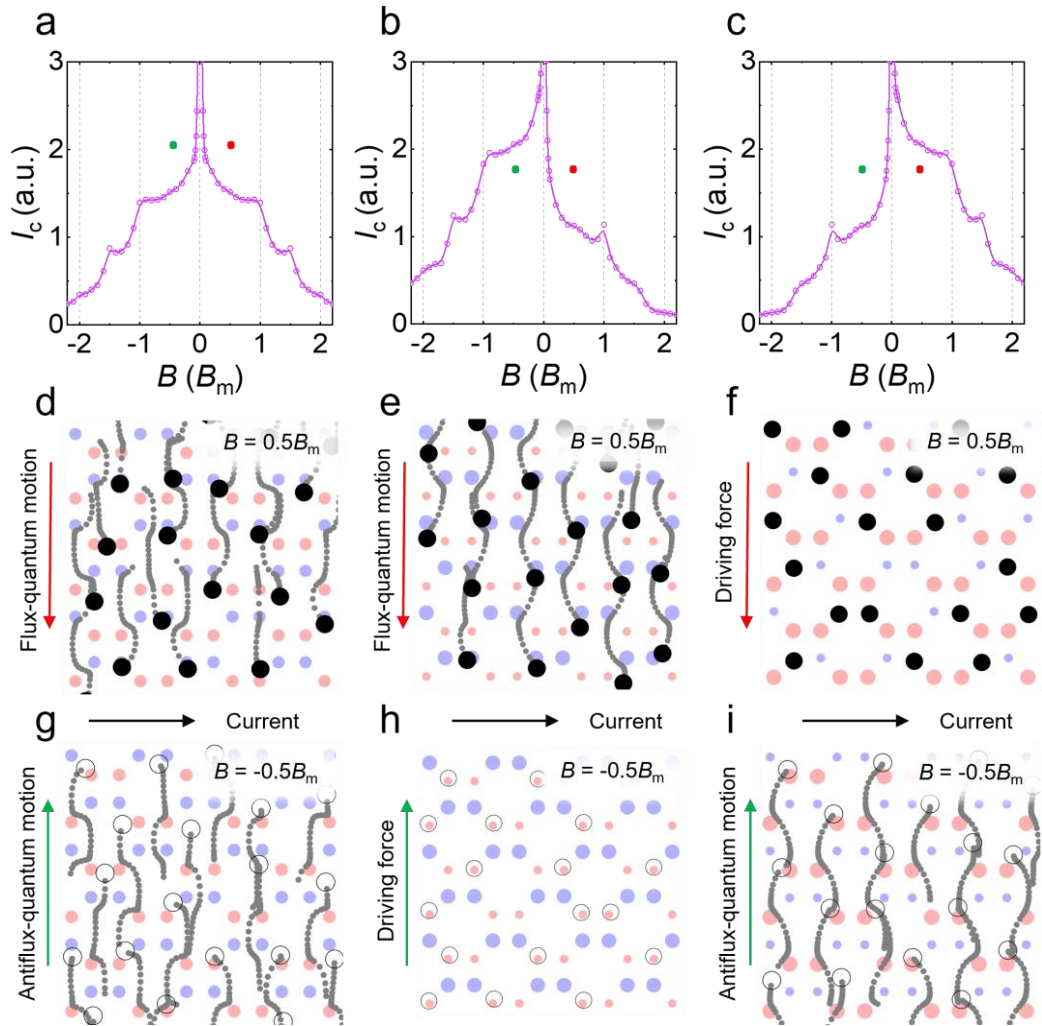

**Fig. 3. Molecular dynamic simulations.** a-c, MD simulations of $I_c \sim B$ curves using the charge patterns in Sections S1-S3, respectively. d-i, Screen shots of vortex motion/trajectories extracted from Videos 1 (S1, d and g), 2 (S2, e and h), and 3 (S3, f and i) and at $B = 0.5B_m$ (d-f) and $B = -0.5B_m$ (g-i), respectively. Red and blue dots are positive and negative magnetic charges, respectively. Black dots and open circles denote vortices and antivortices, respectively. The current and magnetic field values used for simulations are shown by the green and red points in Figures a-c.

The asymmetric nanomagnets in Sections S2 and S3 generate different-sized positive and negative magnetic charges. Both MFM images (Figures 1f and 1g) and micromagnetic simulations (insets of Figures 1f and 1g) indicate that the wider end of each nanomagnet produces a smaller magnetic charge compared to the narrow end. Furthermore, the total magnetic flux value of the positive charge is equal to that of the negative charge from the same nanomagnet. Consequently, the stray field strength of the smaller positive charge in Section S2 is more intense than that of the larger negative charge. Meanwhile, the positive charge attracts antivortices ($B < 0$), while the negative charge attracts vortices ($B > 0$).[30, 31] This results in stronger antivortex pinning under a negative magnetic field compared to vortex pinning under a positive magnetic field. As demonstrated by MD simulations in Video 2 (with corresponding screenshots in Figures 3e and 3h), vortices ($B > 0$) navigate smoothly, whereas the antivortices are immobilized and pinned by the smaller yet stronger positive charges. Similar to the electric nonreciprocal effect in a superconducting diode, the magnetic nonreciprocal effect manifests a magnetic field-driven 'diode' effect, where the system exhibits zero resistance only when the magnetic field is applied in one specific direction.

As mentioned above, magnetic nonreciprocity stems directly from the geometric asymmetry of the nanomagnet. In the case of sample Section S3, the asymmetric nanomagnets are geometrically inverted compared to those in Section S2, leading to the inverted size and field strength of the positive and negative charges. In this scenario, the smaller and more intense negative charges provide stronger pinning to vortices ($B > 0$), while the larger positive charges provide weaker pinning to antivortices ($B < 0$). This facilitates easier motion for antivortices ($B < 0$) compared to vortices ($B > 0$). This inversion results in the polarity of the magnetic nonreciprocal effect being flipped, which is vividly demonstrated in the MD simulations of Video 3 (with corresponding screenshots in Figures 3f and 3i).

While inverting the asymmetric nanomagnet structure can effectively reverse the magnetic nonreciprocal effect, this method is not in situ, as the geometric structure of the nanomagnet cannot be changed post-fabrication of the sample device. However, A-

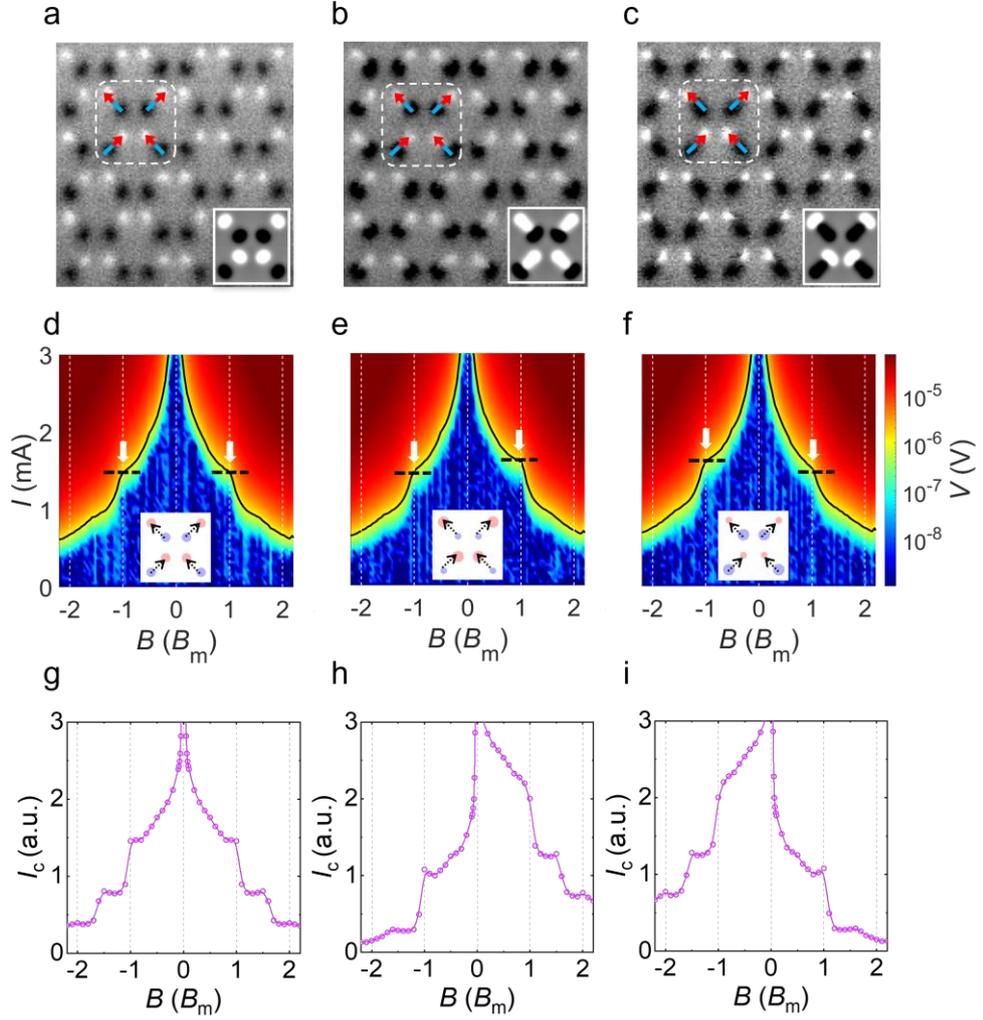

**Fig. 4. In-situ reversible magnetic nonreciprocal effect.** a-c, MFM images of inversely magnetized ASIs in Sections S1-S3, respectively. d-f, Experimental dissipation voltage induced by vortices/antivortices motion, corresponding to Figures 4(a-c). g-i, Corresponding MD simulations of $I_c \sim B$ curves.

SIs provide the advantage of in-situ reconfigurability of their magnetic charge patterns through the application of in-plane polarizing fields, allowing for programmable control of vortex dynamics. MFM images of Figures 4a-4c illustrate the magnetic charge patterns of the inversely magnetized state compared to Figures 1e-1g. When the magnetization state of nanomagnets in ASIs is reversed, the polarities of magnetic charges are switched. In the case of Section S1 with symmetric nanomagnets, given that the size and strength of the positive and negative charges remain consistent, reversing the magnetization of the nanomagnets does not change the relative pinning strengths between vortices and antivortices. The transport results continue to show a magnetically symmetric response, as evidenced by the experimental data in Figure 4d and the simulation data in Figure 4g. Conversely, reversing the magnetization of asymmetric nanomagnets results in inverted sizes and strengths of positive and negative magnetic charges (see Figures 4b and 4c compared to Figures 1f and 1g, respectively), which in turn inverts the relative pinning strengths between vortices and antivortices. For instance, in sample Section S2, the smaller but stronger negative charge leads to a greater pinning strength for vortices, resulting in higher critical currents under a positive magnetic field, as shown in the experimental result in Figure 4e and the MD simulation result in Figure 4h. This process enables an in-situ reversible magnetic nonreciprocal effect. The magnetic nonreciprocity of Section S3 is also inverted, as depicted in Figures 4f and 4i. The inverted magnetic nonreciprocal effects for Sections S2 and S3 are directly observable in Videos 4 (Section S2) and 5 (Section S3), respectively.

We introduced a novel ASI-superconductor heterostructure comprised of asymmetric nanomagnets situated on top of a superconducting film. These asymmetric nanomagnets generate positive and negative magnetic charges of nonidentical sizes and strengths, which provide different pinning forces on vortices and antivortices, leading to their distinct motion behaviors. Under specific currents, this system exhibits zero resistance in only one direction of the magnetic field while becoming resistive in the opposite direction, functioning as a magnetic field-driven superconducting 'diode'. Owing to the reconfigurability of ASIs, the polarity of the magnetic nonreciprocity can be reversed in situ by inverting the magnetization state of the ASI. Further investigation is required to explore the temperature dependence of these magnetic nonreciprocal effects and the in-plane field hysteresis effects. Beyond the square-ASI, our method of employing asymmetric nanomagnets is adaptable to other ASI structures, including but not limited to kagome-ASI,[32, 33] pinwheel (or chiral)-ASI,[31, 34)] and square-charge-ice-ASI.[30, 35] Our research underscores that ASI offers a pinning potential-by-design strategy for modulating superconductor properties and introduces novel functionalities to enhance superconducting electronics.

# Supplementary Information

**Methods**

    **Sample fabrication.** Sample fabrication details can be found in ref. [30] in this work, the ASI array in Section S1 comprises asymmetrically trapezoidal nanomagnets, each sized at 300 nm (length) x 80 nm (width) x 25 nm (thickness), The ASI array in Section S2 comprises asymmetrically trapezoidal nanomagnets, each sized at 300 nm (altitude) x 40 nm (base1) x 120 nm (base2) x 25 nm (thickness), The ASI array in Section S3 features geometrically inverted nanomagnets relative to those in Section S2. The MoGe film has a superconducting transition temperature of 6.6 K, as shown in Supplementary Figure S3.

    **Transport experiments.** Transport measurements were conducted at a temperature of 5.7 K in a triple-axis superconducting vector magnet capable of providing a magnetic field in any desired 3D orientation. See details in ref. [30]

    **Molecular dynamics simulation.** The size of the simulated sample is L×L, where $L = 25.6\lambda$ ($\lambda$ is the London penetration depth), For symmetric nanomagnets, we use $r_p = 0.3$ and $F_p = 0.5$. In the case of asymmetric nanomagnets, the small charges have $r_p = 0.22$, $F_p = 0.93$, while the large charges have $r_p = 0.45$, $F_p = 0.22$. details on the method are described in ref. [30]

    **Micromagnetic simulation.** We use the MuMax3 code to conduct micromagnetic simulations.[36] The material parameters used for permalloy were as follows: saturation magnetization $M_s = 8.6 \times 10^5$ A/m, exchange stiffness $A_{ex} = 13$ pJ, and damping factor α = 0.01. After the magnetization relaxed to a stationary state (ground state), the z component of demagnetizing field $B_{demag}$ and the MFM images at 50 nm above the nanomagnets were extracted for plotting the magnetic charge patterns shown in the manuscript.

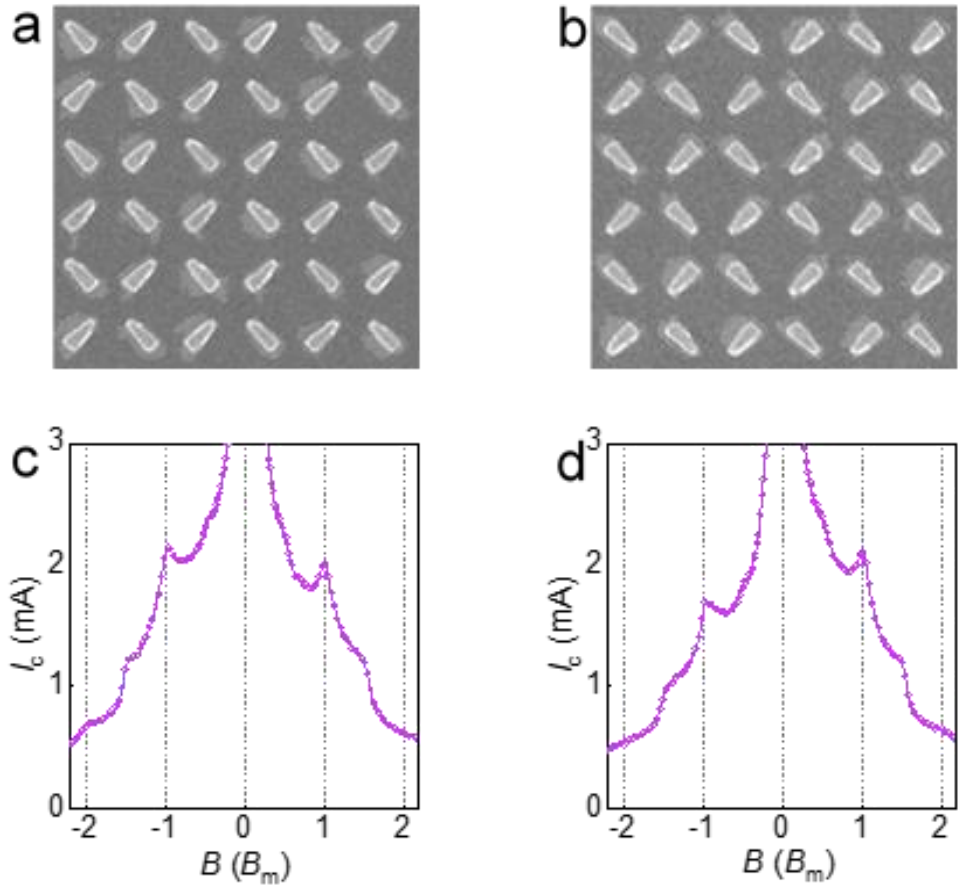

**Fig. S1. Magnetic nonreciprocal effect from another sample.** a and b, SEM images. c and d, field-dependent critical currents for the sample sections corresponding to (a) and (b), respectively.

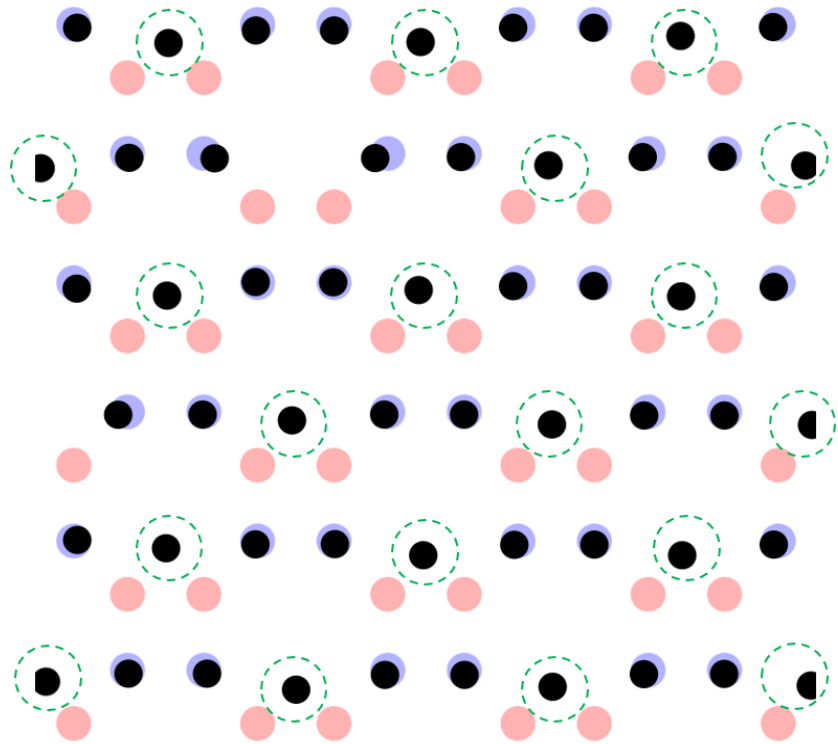

**Fig S2. Vortices distribution at** $B = 1.5\ B_m$. Interstitial vertices are highlighted by green dashed circles.

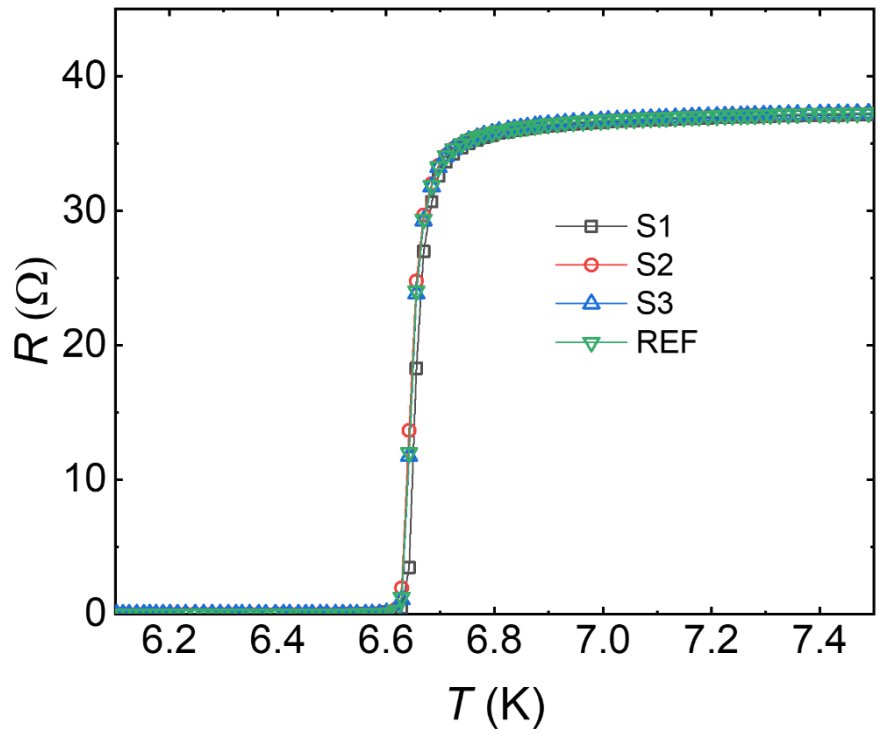

**Fig. S3. Temperature-dependent resistance.** Four curves were obtained from the four sections shown in Figure 1a. They were measured under a current of 100 µA and at zero magnetic field. The critical temperature is about 6.6 K.

**Video 1.** Vortices/antivortices exhibit equivalent motion under positive (vortices) and negative (antivortices) fields, leading to a symmetric transport response. The driving current is set to 2.0. Red and blue dots denote positive and negative magnetic charges, while black dots and open circles represent vortices and antivortices, respectively.

**Video 2.** Vortices (black dots) move smoothly whereas antivortices (open circles) are immobilized and pinned under the same driving current $I$=1.7.

**Video 3.** Antivortices (open circles) move smoothly whereas vortices (black dots) are immobilized and pinned under the same driving current $I$=1.7.

**Video 4.** Reversed vortex/antivortex motion behavior under inverted magnetic charge patterned as compared to Video 2.

**Video 5.** Reversed vortex/antivortex motion behavior under inverted magnetic charge patterned as compared to Video 3.